\documentclass[11pt,dvips]{article}
\usepackage{cite}
\usepackage{graphicx}
\usepackage{psfrag}
\usepackage{url}

\begin{document}
\begin{titlepage}
\title{Phase Transitions in Systems of Interacting Species}
\author{Margarita Ifti \\
Department of Physics, Faculty of Natural Sciences, \\
University of Tirana, Bul. Zogu I, Tirana, Albania \\
Birger Bergersen \\
Department of Physics and Astronomy, \\
University of British Columbia, 6224 Agricultural Road, \\
Vancouver, BC, Canada V6T 1Z1}
\date{}
\maketitle
\normalsize 

\begin{abstract}

We discuss an autocatalytic reaction system: the cyclic competition model
$A_1 + A_2 \rightarrow 2 A_2$, $A_2 + A_3 \rightarrow 2 A_3$, $A_3 + A_4
\rightarrow 2 A_4$, $A_4 + A_1 \rightarrow 2 A_1$), as well as its neutral
counterpart. Migrations are introduced into the model. When stochastic
phenomena are taken into account, a phase transition between a
``fixation'' and a ``neutral'' regime is observed. In the ``fixation''
regime, species $A_1$ and $A_3$ form an alliance against species $A_2$ and
$A_4$, and the final state is one in which one of the symbiotic pairs has
won. The odd--even ``coarse--grained'' systems is mapped onto the 
two--species neutral (Kimura) model. In the ``neutral'' regime, all four 
species survive for long (evolutionary) times. The analytical results are 
checked against computer simulations of the model. The model is generalized 
for $n$ species. Also, a generalized version of the Volterra model is analysed.

\end{abstract}

\end{titlepage}

\section{Introduction}

There is a class of processes in which the competition plays a very 
important role. Examples are ecological, political, epidemiological,
economic, chemical, reaction-diffusion, biological systems. An important
sub-class of those is the cyclic competition systems. In ecology, cases
when variants of a species compete with one-another in a cyclic fashion
have been observed~\cite{reeves72, james91, sinervo96, smith96}. Another
system of interest are cyclic food webs. In politics, different political
parties compete and replace one-another in the helm of power. In the
epidemiological context, examples are diseases which do not leave 
permanent immunity, known otherwise as $SIRS$
(Susceptible-Immune-Recovered-Susceptible) models~\cite{cooke77,
longini80}. Goodwin~\cite{goodwin63} introduced a system of interacting
biochemical metabolic oscillators, which has an autocatalytic feedback
mechanism. Biochemical reactions in a cell support its activities, hence
assuring its very existence. Autocatalytic reactions are an important
class of reactions within a cell. They are reputed to have made possible
the birth and existence of life itself. 

The simplest example of an autocatalytic reaction is the loop of the type
$A_i + A_{i+1} \rightarrow 2 A_{i+1}$, where $i=1, \ldots, k; A_{k+1} =
A_1$. The molecules are in a well-stirred container (the cell), which is
in contact with a reservoir (the outside environment). They can migrate
into and out of the container, to and from the reservoir. In another
(ecological) context, the $\{A_i\}$'s are versions of a biological
species, and in the epidemiological one, states of an individual (e.g. 
susceptible, infected, etc.) In ecological systems it makes sense to also
study the neutral version of this model, in which $A_i + A_{i+1}
\rightarrow 2 A_{i+1}$ or $2A_i$ with equal probability, corresponding to
the Kimura model of neutral genetic drift~\cite{kimura68, kimura69}.

Duty~\cite{tim} has worked on the two allele almost neutral drift model 
with mutations. The almost neutral model with mutations, preserving the 
total number of individuals, has only one degree of freedom, and allows 
one to derive an ``effective potential'' from the Fokker-Planck equation,
obtained by a Kramers-Moyal expansion of the master equation. For small
mutation probabilities, such that $2\mu N \ll 1$, there is extinction of
one species and fixation. The effective potential is almost symmetric
around the centre (where both species are in equal numbers) and the
branches of the effective potential are down. This allows for the system
to quickly slip into a state where only one of the species is present. 
Otherwise, both species coexist forever in the high mutation regime,
i.e. when $2 \mu N \gg 1$. In that regime, the effective potential is
symmetric around the centre point, but with branches upwards, which means
that the centre is a minimum potential point. The system then remains in 
the vicinity of that point for very long times. The effective potential 
``flips'' from ``branches up'' to ``branches down'' at the point where $2 
\mu N =1$. The transition is second-order, and critical behaviour is 
observed.

In two previous studies~\cite{unebirger1, unebirger2} we have considered
an $ABC$ model with cyclic competition/neutral drift and mutations
(migrations) at a constant probability. The system exhibits a critical
transition from a ``fixation'' regime to a ``neutral'' one, in which
biodiversity is preserved over long time. In the ``fixation'' regime, the
number of the $A, B, C$ species oscillates with an amplitude that drifts
with time, until one of the species (and then the second one) goes extinct, 
except for occasional ``bursts'' (which are absent when there is no
migrations in the picture). In the ``diversity'' (or ``neutral'') regime,
the number of the $A, B, C$ versions fluctuates around the centre point,
and there are rare extinctions, but the product $ABC$ remains nonzero
almost always. The survival probability decays exponentially below the
transition point, but the exponent decreases as the mutation (migration) 
probability per particle increases, until it becomes zero at the critical
point. The critical mutation (migration) probability depends on system
size as $1/3 N$, and the models have the same power-law exponent: $-1$. 
There is no qualitative difference between the system with mutations and
that with migrations.

In the present paper we study the system with four or more species, and
show that the above-described picture holds. We show that the cyclic
system is a generalisation of the well-known Lotka-Volterra system, and
that the size-induced transition is present in those systems as well.

\section{The Model}

Our system is an autocatalytic loop of the type $A_i + A_{i+1} \rightarrow
2 A_{i+1}$, where $i=1, \ldots, k; A_{k+1} = A_1$. The molecules are in a
well-stirred container (the cell), which is in contact with a reservoir
(the outside environment). They can migrate into and out of the container,
according to the following rules: a molecule (individual) of species $i$ 
leaves the container at a rate $D \cdot a_i$, and enters it at a rate $D 
\cdot s_i$, where $a_i$ and $s_i$ are its concentrations in the cell and 
the reservoir, respectively. (We will assume for simplicity that the 
reservoir is large enough, so that the exchanges do not perturb it.) The 
rate equations then read:

\begin{equation}\label{eq:rateqm}
\frac {da_i}{dt} = a_{i-1} a_i - a_i a_{i+1} + D (s_i - a_i)
\end{equation}

\noindent In the rate equations approximation, the system size is 
conserved. The above equations~(\ref{eq:rateqm}) have a fixed point, and it
is a stable solution.

The above model relates to the famous Lotka-Volterra~\cite{lotka20,
volterra31} model of interacting populations. A very good, and pleasant to
read, review of the model has been written by Goel, Maitra, and
Montroll~\cite{montroll71}. Volterra described the system of two kinds of
fish (predator-prey) by the pair of equations:

\begin{eqnarray}\label{eq:vol2}
\frac{dN_1}{dt}=\alpha_1 N_1 -\lambda_1 N_1 N_2 \\
\frac{dN_2}{dt}=-\alpha_2 N_2 +\lambda_2 N_1 N_2 \nonumber
\end{eqnarray}

The terms of the form $\lambda_i N_1 N_2$ describe respectively the
depletion of the stock of fish 1 and the enrichment of that of fish 2 from
eating fish 1, and those of the form $\alpha_i N_i$ the evolution of the
species $i$ in absence of the other. Further on, he generalized the pair
of equations (\ref{eq:vol2}) to $n$ species:

\begin{equation}\label{eq:voln}
\frac{dN_i}{dt}=k_i N_i + \beta_i^{-1} \sum_{j=1}^n a_{ij} N_i N_j
\end{equation}

\noindent In absence of other species, the $i$-th species will grow or die
exponentially, depending on the sign of $k_i$. The constants $a_{ij}$ will
be positive, if species $i$ eats $j$, negative, if it is eaten by $j$, and
zero, if they do not interact at all. This leads to $a_{ij}=-a_{ji}$. 

The mean-field (rate equations) behaviour of the system is well-known. It
accepts periodic solutions, as shown by Volterra. Statistical mechanics of
the Volterra system was first constructed by Kerner~\cite{kerner57}. Goel
et al.~\cite{montroll71} studied the existence and stability of the
solutions to the rate equations. They also introduced a random function of
time (noise) in the growth equation, obtaining a Liouville equation. The
model is indeed one of the simplest of nonlinear competition models.

Jain and Krishna~\cite{jain02} use directed graphs to describe
autocatalytic sets. Directed graphs are a set of `nodes' and `links',
where each link is an ordered pair of nodes~\cite{harary69, bangj01}. A
graph with $p$ nodes is specified by its adjacency matrix $C$, defined in
such a way that the $c_{i,j}$ element is one, if the graph contains a link
directed from node $j$ to node $i$ and zero otherwise. For any
non-negative matrix, the Perron-Frobenius theorem~\cite{seneta73,
berman94} states that there exists an eigenvalue which is real and larger
than or equal to all the other eigenvalues in magnitude. It has been
shown~\cite{jain99} that when a graph has a closed loop, the
Perron-Frobenius eigenvalue of its adjacency matrix is exactly 1. The set
of Perron-Frobenius eigenvectors is of interest to us, since it provides
the attractors for the dynamical system whose evolution is described by
the set of differential equations:

\begin{equation}
\dot{a_i} = \sum_{i=1}^s c_{i,j} a_j - \phi a_i \nonumber
\end{equation}

\noindent which is equivalent to the generalized Volterra equation:

\begin{equation}
\dot{a_i} = \sum_{i=1}^s c_{i,j} a_j - a_i \sum_{k,j} c_{kj} a_j
\end{equation}

\noindent when the catalysed reaction is much faster than the one
described by the first term (the spontaneous one). The system will then
converge towards the fixed point that is a Perron-Frobenius eigenvector of
the graph's adjacency matrix. In the case of our cyclic systems, the
Perron-Frobenius eigenvalue is one, and the components of the
corresponding eigenvector are all equal. This means that, in the rate
equations approximation, the symmetric system (with all the rates equal to
one) will approach the centre (all $a_i$'s are equal), and remain there.

A description of the newest techniques of stability analysis for such 
systems can be found in~\cite{albert}, and the references therein.
 
However, the rate equations are just a ``mean field'' approximation; they
only describe the behaviour of the average values of the individual
populations. In the real world, the system is subject to stochastic noise
due to birth and death processes (intrinsic noise), which we take to be
Poisson-distributed. The random nature of these processes need be taken
into consideration, if we want to obtain the correct and complete
behaviour of the system. For that we ought to write the master equation,
and then somehow solve it. Unfortunately, very few master equations are
simple enough to accept analytical solutions. We deal with this situation
by expanding them into a Fokker-Planck equation, which then helps us draw
the necessary information about the behaviour of the system. Another
approach is to simulate the master equation of the system. 

The master equation for, say, the four-species cyclic system is:

\[
\frac{\partial P(A_1,A_2,A_3,A_4,t)}{\partial t} = \frac{1}{N} \left [ ( 
\epsilon_4 \epsilon^{-1}_1 -1) A_1 A_4 + (\epsilon_1 \epsilon^{-1}_2-1) 
A_1 A_2 + \right. \]\begin{equation}\label{eq:mast4}
\left. + (\epsilon_2 \epsilon^{-1}_3 -1) A_2 A_3 +(\epsilon_3 
\epsilon^{-1}_4-1) A_3 A_4 \right ] P(A_1,A_2,A_3,A_4,t)
\end{equation}

\noindent where we have used the ``shift'' operators notation:

\begin{eqnarray}\label{eq:shift}
\epsilon_1 f(A_1,A_2,A_3,A_4) = f(A_1+1,A_2,A_3,A_4) \nonumber \\
\epsilon^{-1}_1 f(A_1,A_2,A_3,A_4) = f(A_1-1,A_2,A_3,A_4)
\end{eqnarray}

\noindent and similarly for the other concentrations.

\section{System-Size Expansion of the Master Equation}

For systems like the one above, where the rate equations have a stable 
solution, the $\Omega$-expansion of van Kampen~\cite{vankampen97} works 
exceptionally well. Previously we have used it to solve a three-species
cyclic and neutral system in presence of mutations and
migrations~\cite{unebirger2}. 

Using the ``shift'' operators notation~(\ref{eq:shift}), the master
equation for the cyclic competition system with migrations reads:

\[
\frac{\partial P( \{ A_i \} ,t)}{\partial t} = \left \{ \frac {1}{N}
\left [(\epsilon_1 \epsilon^{-1}_1 -1) A_1 A_4 + (\epsilon_1 
\epsilon^{-1}_2 -1) A_1 A_2 + \right. \right. \]\[ \left. + (\epsilon_2 
\epsilon^{-1}_3-1) A_2 A_3 +(\epsilon_3 \epsilon^{-1}_4 -1) A_3 
A_4 \right ]+ D \left [(\epsilon_1-1)A_1 + (\epsilon_2-1) A_2 + \right. 
\]\begin{equation}\label{eq:mig}
\left. \left. + (\epsilon_3 -1) A_3 + (\epsilon_4 -1) A_4 +\frac{N}{4} 
(\epsilon^{-1}_1+ \epsilon^{-1}_2 + \epsilon^{-1}_3 + \epsilon^{-1}_4 -4) 
\right ] \right \} P(\{ A_i \} ,t)
\end{equation}

The idea of the van Kampen expansion~\cite{vankampen97} is to split the 
variables of the problem into a non-fluctuating part, and a fluctuating 
one, i.e. deal separately with the mean-field solutions and the 
fluctuations (which are taken to be of the order $\sqrt N$). In this
approach, the numbers of the individual populations would be written as:

\begin{equation}\label{eq:splitv}
A_i=N \phi_i + \sqrt N x_i
\end{equation}

\noindent Here the $\phi_i$ are the steady-state (non-fluctuating) 
concentrations of the $i$-th  species respectively (which only depend on
time), and the $x_i$ are the fluctuations (proportional to the square root
of system size). Then the probability distribution $P(A_i,t)$ is
transformed into $\Pi(\{x_i\},t)$, and the following relations are true:

\[ \Pi = N^2 P(N \{\phi_i + \sqrt N x_i\},t) \]
\begin{equation}\label{eq:vartrans}
\frac {\partial P}{\partial t} = \frac {1}{N^2} \frac {\partial
\Pi}{\partial t} - \frac{1}{N} \sum \frac {d \phi_i}{d t} \frac 
{\partial \Pi}{\partial x_i} \nonumber
\end{equation}

\noindent and

\begin{eqnarray}\label{eq:epsexp}
\epsilon_i = 1+ \frac{1}{\sqrt N} \frac {\partial}{\partial x_i} + \frac
{1}{2N} \frac {\partial^2}{\partial {x_i}^2} + \ldots \\
{\epsilon_i}^{-1} = 1- \frac{1}{\sqrt N} \frac {\partial}{\partial x_i} +
\frac {1}{2N} \frac {\partial^2}{\partial {x_i}^2} + \ldots \nonumber
\end{eqnarray}

Next we substitute everything into the master equation, leave only the 
term $\partial \Pi / \partial t$ on the left hand side, and group the 
right hand side terms according to powers of $\sqrt N$. The first term is
of order $N^{1/2}$, and it must be equal to zero, for an expansion in
terms of $N^{1/2}$ to make sense. That term reproduces the rate equations
in terms of the concentrations $\phi_i$, with steady state solution
$\phi_i=1/4$.

The terms of order $N^0$ give a linear Fokker-Planck equation of the form:

\begin{equation}\label{eq:fpsurv}
\frac {\partial \Pi}{\partial t} =\sum [-A_{ik} \frac {\partial}{\partial
x_i} (x_k \Pi) + \frac{1}{2} B_{ik} \frac {\partial^2 \Pi}{\partial x_i
\partial x_k}]
\end{equation}

\noindent where the A-matrix for the cyclic system is:

\[ \left(
\begin{array}{cccc}\label{eq:amatmig}
\phi_4-\phi_2- D & - \phi_1& 0 & \phi_1 \\
\phi_2 & \phi_1-\phi_3- D & - \phi_2 & 0 \\
0 &- \phi_3 & \phi_2 - \phi_4 - D & -\phi_3 \\
- \phi_4 & 0 & \phi_4 & \phi_3-\phi_1- D \end{array}
\right) \]

\noindent and for the neutral system:

\[ \left(
\begin{array}{cccc}\label{eq:anetmig}
- D & 0 & 0 & 0 \\
0 & - D & 0 & 0 \\
0 & 0 & - D & 0 \\
0 & 0 & 0 & - D \end{array}
\right) \]

\noindent The B-matrix is the same for both systems. Its diagonal elements 
are $B_{ii}=D(s_i+\phi_i)+\phi_i(\phi_{i-1}+\phi_{i+1})$, and the 
off-diagonal ones: $B_{ij}=-\phi_i \phi_j$. The Fokker-Planck equation 
obtained this way is linear, and the coefficients depend on time through 
$\phi_i$. We are interested in fluctuations around the steady state. This
approximation is otherwise known as ``linear noise approximation''. The
solution is known to be a Gaussian; the problem represents itself as an
Ornstein-Uhlenbeck process. For our purposes, it suffices to determine the
first and second moments of the fluctuations. Following van
Kampen~\cite{vankampen97}, we can multiply the Fokker-Planck equation by
$x_i$ and integrate by parts to get:

\begin{equation}\label{eq:avgfl}
\frac {d \left <x_i \right >}{dt} = \sum_j A_{ij} \left <x_j \right >
\end{equation}

\noindent For simplicity we can assume that all the concentrations in the 
reservoir are equal: $s_i=s=1/4$. The eigenvalues of the $A$ matrix are 
$-D$ (doubly-degenerate), $-D \pm i \phi \sqrt 2$ for the cyclic system, 
and $-D$ (quadruply degenerate) for the neutral system. Here $\phi$ is the 
steady state value of the concentrations (we have dropped the index, since 
they are all the same). The negativity of the eigenvalues guarantees the 
stability of the zero solutions to the first moments equations. Hence, the 
average of the fluctuations decays to zero and remains zero. The equations 
for the second moments can be obtained similarly:

\begin{equation}\label{eq:varflu}
\frac {d \left <x_i x_j \right >}{dt} = \sum_k A_{ik} \left <x_k x_j 
\right > + \sum_k A_{jk} \left <x_i x_k \right > + B_{ij}
\end{equation}

They depend on time through $\phi_i$'s, which we again substitute by their
steady state value, since we are interested in the fluctuations around
that state. Also, by symmetry, all the diagonal terms $\left <x_i^2 
\right >$ are equal, as well as off-diagonal terms (correlations) $\left 
<x_i x_j \right >$. They depend on the migration probability $D$ alone. The 
steady state solutions for the diagonal terms (and also for the variances, 
since the mean values are zero), are as follows:

\begin{equation}\label{eq:solmig}
\left <x_i^2 \right > = \frac {(2 \phi^2 +D(\phi + s))}{2D}
\end{equation}

\noindent where $s$ is the concentration of any of the species in the 
reservoir. The number of individual species will fluctuate around $N \cdot
\phi_s$ where $\phi_s$ is the steady-state concentration. All the species
survive forever. This way, (sufficient) migrations into and out of the
container maintain diversity in the system.

It is important to point out that from the expression for the A-matrix
above, we can tell that when migrations are absent ($D=0$), the (real part
of all) eigenvalues of that matrix become zero, and the equation is of the
diffusion type~\cite{vankampen97}. In that case, the rate equations
suggest that the non-fluctuating part will remain constant (at the
centre). However, small deviations give rise to large differences, and one
would expect the fluctuations to grow, rather than remain limited. The
separation into a macroscopic part and small fluctuations is no longer
meaningful. In this case, after a transient period (of the order
$N^{1/2}$), one would expect $P$ to be a smooth function of the
concentrations, and expand in powers of $N$, rather than $N^{1/2}$. This
is otherwise known as Kramers-Moyal expansion, and we have employed it for
the three-species systems~\cite{unebirger1, unebirger2}, in very good
agreement with the simulations results.

Furthermore, the only difference between the three and four species 
systems expansion is the appearance of an ``off-diagonal'' of zeroes, in
positions $(i,j)$ for which the species $A_i$ and $A_j$ do not react. This
means that the above algebra will remain exactly valid when there are more
than four species in the system. Hence, our results will hold for any
number of species, and any reasonable system size.

If there is only migrations into and out of the container (i.e. no cyclic
reactions), the system remains near the centre point, and the product of
the concentrations remains considerably above zero; in other words, all
species are present in the system at (almost) all times. When both the
cyclic/neutral mechanism and migrations are present, one can occasionally
observe temporary extinctions. Since the boundary is not absorbing,
occasional migrations will return the system to the state with maximal
symmetry (diversity) where all species coexist. The migrations then manage
to keep the system maximally disordered, since they are stronger than the
fluctuations (which try to drive the system toward the boundary, i.e. 
fixation, and keep it there). The migration rate acts then as some sort of
``temperature'', and decreasing the migration rate would be analogous to
annealing the system.

\section{The Transition Region}

In two papers, Togashi and Kaneko~\cite{togashi01, togashi02} focus on the 
four-species autocatalytic system with a very small number of molecules. 
Their work covers many aspects of a transition, which they baptize 
``discreteness-induced''. The idea of their work is that, when the number 
of molecules in the system is very small, a transition from the state where 
the numbers of molecules of the individual types are Gaussian-distributed, 
as derived above, to one in which the $A_1$ and $A_3$ species form an 
alliance against species $A_2$ and $A_4$ is observed. In their work they 
keep the diffusion rate $D$ constant, and vary the system size. As the 
total number of molecules (system size) decreases and goes through a 
certain value (which coincides with $1/D$, the inverse of migration rate), 
they observe a transition into the broken-symmetry state. One of the 
symbiotic pairs eventually wins, but, since there are migrations into and 
out of the container happening all the time, the introduction of a member 
of the opposite symbiotic pair may bring the victory of that initially 
(almost) non-existent pair, making the pair that previously was the 
``winner of the hour'' disappear, and so on. The probability distribution 
of the number $z=(x_1+x_3)-(x_2+x_4)$ (corresponding to our $(a_1+a_3) - 
(a_2+a_4)$) is shown in Fig.~2 of their article~\cite{togashi01}, and the 
formation of the peaks that indicate the (temporary) victory of one such 
symbiotic pair over the other as the system size goes through $1/D$ is 
quite pronounced. In Fig.~4 of their PRL article~\cite{togashi01} Togashi 
and Kaneko show a plot of the rate of residence of the 1-3 (or 2-4) rich 
state (i.e. the state in which one of the pairs ``rules'') as a function of 
the product $DV$ ($V$ is system size, the parameter we call $N$). The rate 
of residence of the symmetry-broken state clearly goes to zero, as $DV 
\rightarrow 1$.

Looking at the expression for the variance of the fluctuations 
above~(\ref{eq:solmig}), one can observe that when the migration
probabilities per particle (migration rate) approach zero, the variance of
the concentrations of individual populations is of the order $\frac {(2
\phi^2 + D (\phi + s))}{2D}$, and it becomes of order 1 (i.e. the order of
macroscopic concentrations,) when $D \sim 1/N$ (here we are using the
values of parameters as chosen by Togashi and Kaneko~\cite{togashi01}, who
assume all $\phi_i = s_i =1$. However, it works exactly the same way with 
our steady state values $\phi_i = s_i =1/4$). This gives us the critical 
value for the migration probability. The critical $D$ thus obtained is in 
excellent agreement with the one Togashi and Kaneko observe and show in 
Fig.~2 of their letter~\cite{togashi01}, when they see the system go 
through a transition for $V(N)=1/D$. Also, this verifies the other result 
of their simulations, shown in Fig.~4 of their letter~\cite{togashi01}, 
where the rate of residence of the symmetry-broken state approaches zero, 
as $DV \rightarrow 1$. In our work on three-species 
systems~\cite{unebirger1, unebirger2} we have obtained similar results 
using van Kampen expansion, and verified them in the diffusive (fixation) 
regime, using a Kramers-Moyal expansion, having imposed the condition that 
the system be near a critical transition, i.e. the smallest eigenvalue of 
the Fokker-Planck equation be zero. The agreement between the two is 
excellent.

\section{Steady-State Solutions of the Rate Equations and Simulations
Results in the Fixation Regime}

It is important to know the behaviour of the system, when it is closed,
i.e. no migrations into and out of the container are possible, since the
final state in the fixation (broken-symmetry) regime is the same. The rate
equations for this system, written for the concentrations of individual
species, will be:

\begin{equation}\label{eq:rateq}
\frac {da_i}{dt} = a_{i-1} a_i - a_i a_{i+1}
\end{equation}

\noindent and similarly for the other species. If we consider the case
when $A_i + A_{i+1} \rightarrow 2A_i$ or $2 A_{i+1}$ with equal 
probability, the right-hand side of the rate equations will be identically 
zero.

One can directly find the steady-state solutions of the rate 
equations~(\ref{eq:rateq}), i.e. the concentrations that reduce the 
left-hand side to zero. For any number of species in the system, the 
centre, where all the concentrations are equal, is a solution of the rate 
equations~(\ref{eq:rateq}). When three species are present, the other 
solutions are those for which one of the species is alive, and the others
are extinct. In the case of four and five species, the solutions are of
the form $a_i=a_{i+2}$ (modulo 4 or 5, depending on the number of species
in the system) with all the other concentrations equal to zero. The system
with six species, except for the centre, has two other kinds of sets of
solutions: one with $a_i=a_{i+2}=a_{i+4}$ (modulo 6) and all the others
zero (i.e. three species alive and three dead), and one set with
$a_i=a_{i+3}$ (modulo 6) with the rest of concentrations equal to zero 
(i.e. two ``opposite'' species alive and four dead). For any numbers of
species in the system, the product of concentrations is conserved, as well
as the total number of individuals. This means that any trajectory with
the product of concentrations constant will be a neutrally stable one. In
the case of the neutral systems, the right-hand side is identically zero,
which means that any state is a stable one.

We simulated copies of the four- or more-species cyclic system, for 
different system sizes. These simulations started with equal individual 
populations of $A_1$, $A_2$, $A_3$, and $A_4$ (i.e. the centre). We 
generated times for the next possible reaction event with exponential 
distribution as $- \frac {\ln (rn)}{rate}$, where the rate of the cyclic
or migrations as in the master equation above~(\ref{eq:mast4}) is
substituted. (Here $rn$ is a random variable with uniform distribution in
$[0,1]$. This way we get Poisson distribution for the events, i.e. really 
independent events~\cite{gibson}.) The reaction which occurs first is then 
picked and the system is updated accordingly. The process is repeated for 
a large number of events. We observed the symmetry-breaking phenomenon for 
any system size, no matter how large it is. 

\begin{figure}
\includegraphics{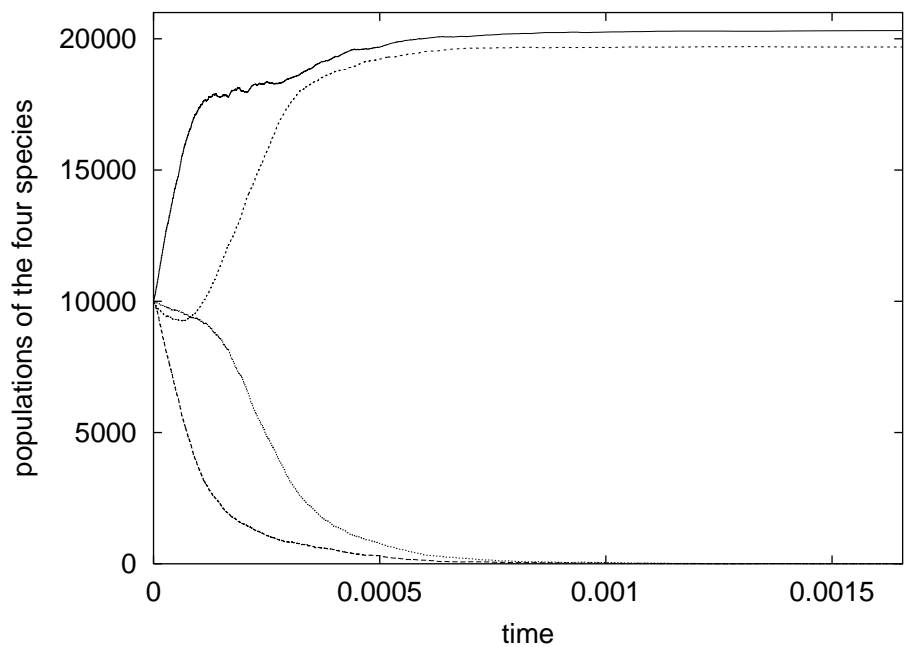}
\caption{The time series for the number of $A_1$, $A_2$, $A_3$, and $A_4$
species in the ``fixation'' regime (here $D=0$, and system size $N=40~000$). 
In this particular realization the $A_1 A_3$ symbiotic pair wins over the
$A_2 A_4$ one.}
\label{fig:tsabcd}
\end{figure}

\begin{figure}
\includegraphics{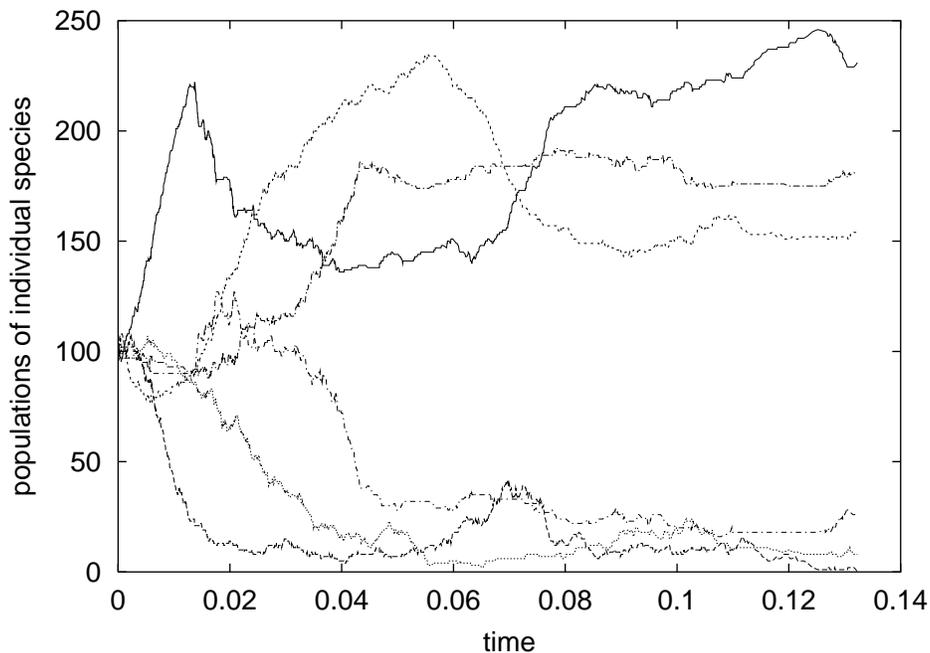}
\caption{The time series for the six-species system ($N=600$). All the 
copies result in a state in which three species (rather than two) survive.}
\label{fig:tsabcdef}
\end{figure}

In Fig.~\ref{fig:tsabcd} we show the time series for the four-species
cyclic system, for $N=40~000$. Contrary to the prediction from the
solutions to the rate equations, for the six-species systems, out of two
thousand realisations of the system, we never obtained a final state with
only two species present. All the copies of the system we simulated ended
up with three species present, as the time series shown in
Fig.~\ref{fig:tsabcdef}. In all the time series, time is measured in units
of the system size $N$.

\begin{figure}
\includegraphics{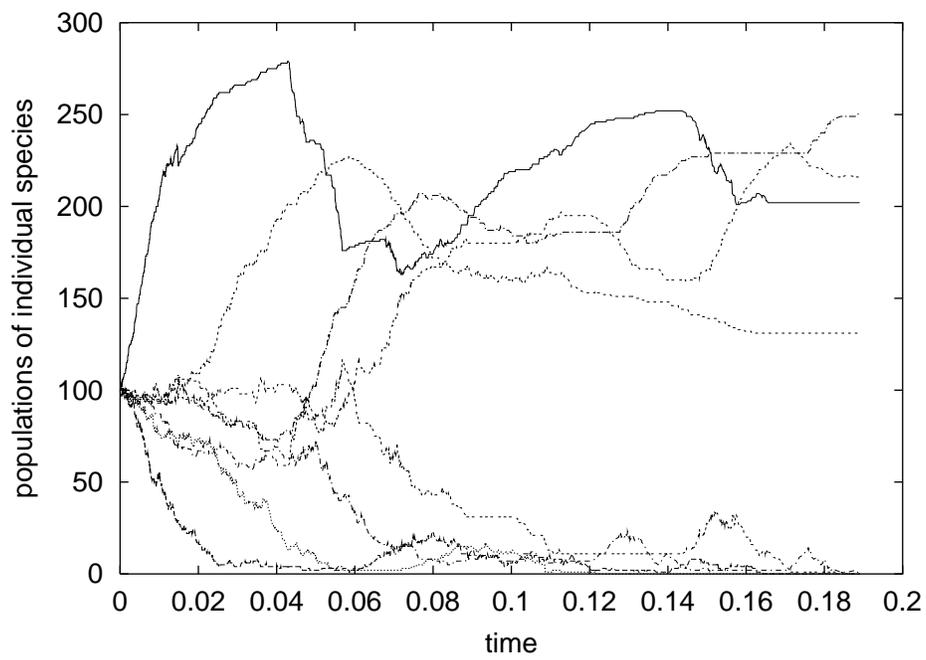}
\caption{The time series for the eight-species system ($N=800$). The
history of the system ends in a state in which four species survive.}
\label{fig:ts8sp}
\end{figure}

When more species are present in the system, the picture gets more 
complicated. However, the general feature is that the system experiences
``fluctuations-generated forces'', which push the system towards the
boundary. There the history of the system ends, since the boundary is
absorbing. When the number of species is odd, the population sizes of
individual species oscillate with an amplitude that drifts with time,
until they reach the boundary, one after another. On the other hand, when
the number of species is even, ``alliances'' do not take long to form, and
the history of the system always ends with half of the species winning
over the other half. A typical time series for an eight species system is
shown in Fig.~\ref{fig:ts8sp}. The fate of the system remains similar
throughout the ``fixation'' regime, i.e. when the diffusion rate $D <
1/N$. In that regime, the system exhibits diffusive behaviour toward the
boundary, and the migrations in and out are not frequent enough to bring
it back to the centre.

The fluctuations push the neutral system towards the boundary, too, but 
the history of the neutral systems ends either with only one species 
present, or with disconnected species, in the sense that they have no 
reason to coexist in symbiosis, since they have no common enemies. On the
other hand, the history of the cyclic systems ends with two or more 
species present. The surviving species do not react with one-another 
directly, but have common ``enemies''. This is quite reminiscent of 
symbiosis (an alga + a fungus = a lichen, or a tree and mushrooms live 
together and help each-other fight common enemies), and yet another 
example of competition giving rise to cooperation.

As a final remark regarding the even-number cyclic system, it is useful to 
try and ``coarse-grain'' its dynamics into two species: odd- and 
even-numbered. For simplicity, let us consider the four-species system. In 
this case:

\[A_{odd}=A_1+A_3 \]\begin{equation}
A_{even}=A_2+A_4
\end{equation}

\noindent and the master equation will transform as follows:

\[\frac{\partial P(A_{odd},A_{even},t)}{\partial t}= \left \{ \frac{2}{N} 
\left [ (\epsilon_{even} \epsilon^{-1}_{odd}-1)+(\epsilon_{odd} 
\epsilon^{-1}_{even}-1) \right ]A_{odd} A_{even} + 
\right.\]\begin{equation}\label{eq:neutral2}
\left. +D \left [(\epsilon_{odd}-1)A_{odd}+ (\epsilon_{even}-1)A_{even} 
+\frac{N}{4} (\epsilon^{-1}_{odd} \epsilon^{-1}_{even}-2) \right ] \right 
\}P(A_{odd},A_{even},t)
\end{equation}

\begin{figure}
\includegraphics{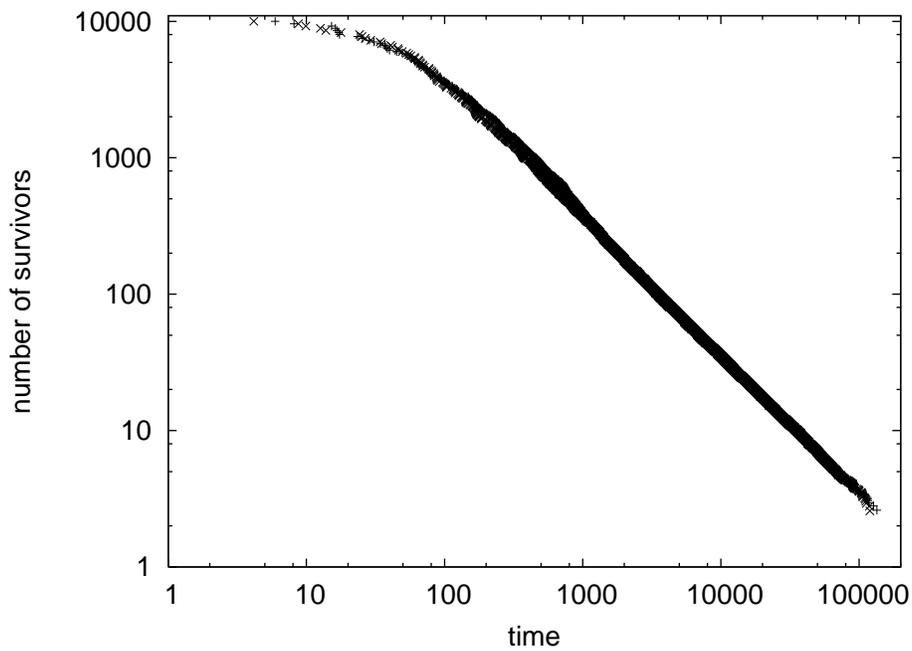}
\caption{Number of survivors vs. time in log--log axes, for the 
``coarse-grained'' odd--even species, system sizes $N=4~000$ and 
$N=20~000$. It maps onto the two-species neutral drift. The slope of 
the line is $-1.03 \pm 0.05$ for $N=4~000$, and $-1.006 \pm 0.009$ for 
$N=20~000$.}
\label{fig:kimura}
\end{figure}

This is exactly the master equation for Kimura's two-species neutral drift 
system~\cite{kimura68, kimura69}, if we recall that the odd--even system 
size is $N'=N/2$. Duty has shown that this system is equivalent to a 
branching process~\cite{resnick} with selection coefficient $s=0$~\cite{tim}.
It is critical, and the extinction times scale as $\sqrt N$, and the 
probability of survival decays with time as $t^{-1}$, as verified from 
simulations (see Fig.~\ref{fig:kimura}).

It is easy to see that for systems with more species (but always an even 
number of them) we will obtain the same ``coarse--grained'' master equation.

\section{The Case of Generalized Volterra Equations}

Now let us go back to the Volterra equations~(\ref{eq:voln}). They can be
generalised further, resulting with the following rate equations:

\begin{equation}\label{eq:gvoln}
\frac{dN_i}{dt}=\sum_j k_{ij} N_j + \frac{1}{N} \sum_{j=1}^n a_{ij} N_i
N_j \equiv f_i(N_1, N_2, \ldots, N_n)
\end{equation}

\noindent where $n$ is the number of species in the system. The diagonal 
elements in both sums correspond to a Malthus-Verhulst equation. The first 
sum now contains off-diagonal elements, which correspond to mutations of 
individuals from species $i$ to species $j$ (or the other way around, 
depending on the sign of the rate $k_{ij}$. These mutation rates will be 
very small.) Hence, the $k_{ij}$'s and $k_{ji}$ have opposite signs (and 
also $a_{ij}$ and $a_{ji}$). The terms of the form $a_{ij} N_i N_j$ 
describe what happens when an $i$ individual runs into a $j$ individual: 
when $a_{ij}>0$ the $i$ individual ``eats'' a $j$ individual and 
reproduces, and the other way around when $a_{ij} < 0$. The master equation 
for the generalised Volterra system, written in terms of the ``shift'' 
operators~(\ref{eq:shift}):

\[
\frac{\partial P(\{N_i\},t)}{\partial t} = \sum_i \left \{ \beta_{ii}
(\epsilon^{-1}_i-1) N_i + \sum_{j \neq i} \beta_{ij} (\epsilon_j
\epsilon^{-1}_i -1) N_j + \right. \]\begin{equation}\label{eq:mastn}
\left. + \frac {\alpha_{ii}}{N} (\epsilon^{-1}_i-1) N_i^2 + \sum_{j \neq i} 
\frac {\alpha_{ij}}{N} (\epsilon_j \epsilon^{-1}_i -1)N_iN_j \right \} 
P(\{N_i\},t)
\end{equation}

\noindent where the way the master equation is written imposes that the 
coefficients $\beta_{ij}$ be chosen as equal to $k_{ij}$ if $k_{ij} > 0$, 
and zero otherwise. Similarly, $\alpha_{ij} = a_{ij}$ if $a_{ij} > 0$, and 
zero otherwise. 

Now we are ready to attempt the van Kampen expansion of the above master 
equation. As before, the variables $N_i$ are split into a non-fluctuating 
and a fluctuating part, as in Eqs.~(\ref{eq:splitv}), which transforms the 
probability density function similarly to~(\ref{eq:vartrans}). After 
substituting everything into the master equation, we obtain a 
Fokker-Planck equation as in~(\ref{eq:fpsurv}), where the A and B 
matrices have dimensions $n \times n$. The diagonal elements of the 
A-matrix will be given by $A_{ii}=\beta_{ii}-\sum_{j \neq i} \beta_{ji} + 
\sum_j ( \alpha_{ij} - \alpha_{ji} ) \phi_j$, and the off-diagonal ones by 
$A_{ij} = \beta_{ij} - ( \alpha_{ij} - \alpha_{ji} ) \phi_i$. The 
existence and stability of the solutions to the Volterra 
equations~(\ref{eq:voln}) has been object of extensive studies. When the 
system is stable, the A-matrix will have negative eigenvalues, which, using 
the equation~(\ref{eq:avgfl}), will mean that the average value of the 
fluctuations will decay to zero. Here we are in luck, since, after a 
careful inspection, we can see that its elements are those of the Jacobian

\begin{displaymath} {\bf J}= \left (\begin{tabular}{cccc}
$\frac{\partial f_1}{\partial N_1}$ & $\frac{\partial f_1}{\partial N_2}$ & 
$\ldots$ & $\frac{\partial f_1}{\partial N_n}$ \\
$\vdots$ & \ & \ & \ \\
$\frac{\partial f_n}{\partial N_1}$ & $\frac{\partial f_n}{\partial N_2}$ & 
$\ldots$ & $\frac{\partial f_n}{\partial N_n}$ \end{tabular} \right )
\end{displaymath}

\

\noindent calculated at the fixed point. If the eigenvalues of this 
Jacobian are negative, the fixed point is stable. The stability of the 
solution can be determined from the {\it Routh-Hurwitz criteria}. The 
interested reader can find a treatment of these and further references in 
Chapter 6 of the book by Pielou~\cite{pielou69}. However, the application 
of Routh--Hurwitz criteria can be tedious and cumbersome, especially when 
dealing with large systems. The technique of {\it qualitative stability} 
helps enormously. It was initiated by the economists Quirk and 
Rupert~\cite{quirk65}, and applied to ecological systems by 
May~\cite{may73}, Levins~\cite{levins74}, and Jeffries~\cite{jeffries74}. 
See Chapter 6 of the book by Edelstein--Keshet~\cite{keshet} for an 
extensive review of this subject matter.

The B-matrix has diagonal elements of the form $B_{ii}=\beta_{ii} \phi_i + 
\sum_{j \neq i} (\beta_{ij} \phi_j + \beta_{ji} \phi_i) + \alpha_{ii} 
\phi_i^2 + \sum_{j \neq i} (\alpha_{ij} + \alpha_{ji}) \phi_i \phi_j$ and 
off-diagonal ones: $B_{ij}= -\beta_{ij} \phi_j - \beta_{ji} \phi_i - 
(\alpha_{ij}+\alpha_{ji}) \phi_i \phi_j$. They are linear in coefficients 
$\beta_{ij}$ and $\alpha_{ij}$, which means that the equations for the 
second moments of the fluctuations (obtained from their general 
expression~(\ref{eq:varflu})) will also be linear in the rate 
coefficients. This will lead to a size-induced transition as the one 
calculated for the autocatalytic loops, and observed in simulations by us 
for the three-species systems~\cite{unebirger2}, and Togashi and Kaneko 
for four-species systems~\cite{togashi01, togashi02}.

\section{Conclusions}

The transition observed by Togashi and Kaneko~\cite{togashi01, togashi02}
in the four-species autocatalytic loops is not specific to systems with a
very small number of particles/individuals. It is the same critical
transition we have previously~\cite{unebirger2} observed for three-species
systems. This transition corresponds to a crossover from a fully symmetric
(``neutral'') state in the high-migration regime, to a ``fixation'' state
for low-migration rates, in which the symmetry is broken in favour of one
or more species. The ``fixation'' regime in the four-species system
exhibits a symbiosis effect, when species $A_1$ and $A_3$ join their
efforts against species $A_2$ and $A_4$, and the final state of the system
is one in which one of the pairs has completely ``eaten up'' the
other. This transition is present for any finite system size. In the
high-migration regime the system allows for a linear noise approximation,
exhibiting itself as an Ornstein-Uhlenbeck process. 

Since the system size is always finite (no matter how large it is), there
is a value of the migration probability per particle (or diffusion
rate) for which the fluctuations of the concentrations become of order
one, and the system undergoes a critical transition. The critical
diffusion rate varies with system size as $1/N$, and the product $DN \sim
1$, i.e. the number of migrants per unit time, necessary for the symmetry
to be preserved in the system (all species to survive) is of the order
1. This result is a bit counterintuitive, since it does not depend on
system size. The analytical calculations are in excellent agreement with
the simulation results, obtained for three- and four-species
systems~\cite{unebirger2, togashi01}. Those analytical calculations
suggest that all the loop-like autocatalytic systems will exhibit the same
critical transition. The form of the equations for the moments of the
fluctuations suggests that the generalised Lotka-Volterra systems will
also exhibit a similar transition.

This situation, known as diffusion-limited reaction, has manifested itself
and been observed in physical systems low dimensions, when diffusion is
not efficient in mixing the reactants. A physical example is the
Ovchinnikov-Zeldovich segregation phenomenon~\cite{zeldovic78}.

In such a situation the hope is that sufficient migration between habitat 
patches will save the system from extinction. Abta and Shnerb~\cite{abta} 
show that the systems in which only the predator is allowed to migrate (such 
as herbivore--plant or parasite insect--plant systems, like the Prickly Pear 
cactus and the moth {\it Cactoblastis cactorum} in Eastern Australia) may 
support oscillations in noisy environment. Much work is yet to be done in 
the future.

\end{document}